%% file: main.tex
\documentclass[hidelinks,conference,10pt]{IEEEtran}
\IEEEoverridecommandlockouts
\usepackage{listings}
\usepackage{lmodern}  
\usepackage{amsmath}  
\usepackage{xcolor}   
\usepackage{balance}

\input{includes/includes}
\input{includes/macros}

\input{includes/pygments}

\usepackage{cite}
\usepackage{amsmath,amssymb,amsfonts}
\usepackage{algorithmic}
\usepackage{graphicx}
\usepackage{textcomp}
\usepackage{xcolor}
\def\BibTeX{{\rm B\kern-.05em{\sc i\kern-.025em b}\kern-.08em
    T\kern-.1667em\lower.7ex\hbox{E}\kern-.125emX}}

\IEEEtriggeratref{25}

\usepackage[switch]{lineno}
\usepackage{hyperref}

\begin{document}


\title{On ML-Based Program Translation: \\ Perils and Promises}

\author{
    \IEEEauthorblockN{Aniketh Malyala$^\dagger$~\thanks{$^\dagger$ Equal contribution.}}
    \IEEEauthorblockA{
        {Silver Creek High School} \\
        San Jose, CA, USA \\
        \href{aniketh.malyala@gmail.com}{aniketh.malyala@gmail.com}
    }
    \and
    \IEEEauthorblockN{Katelyn Zhou$^\dagger$}
    \IEEEauthorblockA{
        {Silver Creek High School} \\
        San Jose, CA, USA \\
        \href{zhoukatelyn@gmail.com}{zhoukatelyn@gmail.com}
    }
    \and
    \IEEEauthorblockN{Baishakhi Ray}
    \IEEEauthorblockA{
        {Columbia University}\\
        New York, NY, USA \\
        \href{rayb@cs.columbia.edu}{rayb@cs.columbia.edu}
    }
    \and
    \IEEEauthorblockN{Saikat Chakraborty}
    \IEEEauthorblockA{
        {Microsoft Research} \\
        Redmond, WA, USA \\
        \href{saikatc@microsoft.com}{saikatc@microsoft.com}
    }
}

\maketitle
\input{body/0.abstract}


\input{body/1.introduction}
\input{body/4.method}

\input{body/5.results}

\input{body/6.related}
\input{body/7.future-work}

\section*{Acknowledgement}
This work is supported in part by NSF grants SHF-2107405, SHF-1845893, IIS-2040961, IBM, and VMWare. 
Any opinions, findings, conclusions, or recommendations expressed herein are those of the authors and do not necessarily reflect those of the US Government, NSF, IBM or VMWare.

\bibliographystyle{IEEEtran}
\bibliography{main}

\end{document}

%% file: includes/includes.tex
\usepackage{hyperref}
\usepackage{boldline}
\usepackage{color,colortbl}
\usepackage{bigstrut}
\usepackage[ruled]{algorithm2e}
\usepackage{amsmath,amsfonts}
\usepackage{amsmath}
\usepackage{array}
\usepackage{algorithmic}
\usepackage{balance}
\usepackage{blindtext}
\usepackage{booktabs}
\usepackage{caption}
\usepackage{hyperref}
\usepackage[noabbrev]{cleveref}
\usepackage{comment}
\usepackage{enumitem}
\usepackage{eqparbox}
\usepackage{fancybox}
\usepackage{fancyvrb}
\usepackage{framed}
\usepackage{graphicx}
\usepackage{ifthen}
\usepackage{listings}
\usepackage{mathrsfs}
\usepackage{mdwmath}
\usepackage{mdwtab}
\usepackage{multirow}
\usepackage{pifont}
\usepackage{stfloats}
\usepackage{textcomp}
\usepackage{url}
\usepackage{xspace}
\usepackage[normalem]{ulem}
\usepackage[framemethod=TikZ]{mdframed}
\usepackage{caption}
\usepackage{subcaption}
\usepackage{tabularx}
\usepackage{amsthm}

\newcounter{theorem}
\theoremstyle{definition}

\usepackage{tcolorbox}
\tcbuselibrary{listings,skins}
\usepackage{mathtools}
\usepackage{blindtext}
\usepackage{lstautogobble}
\usepackage{tikz}
\usepackage{lipsum}

\DeclareGraphicsExtensions{.pdf,.jpeg,.png}

\lstdefinestyle{customc}{
  belowcaptionskip=1\baselineskip,
  xleftmargin=17pt,
  xrightmargin=3pt,
  language=C,
  showstringspaces=false,
  basicstyle=\scriptsize\ttfamily,
  keywordstyle=\bfseries\color{purple!40!black},
  commentstyle=\itshape\color{blue},
  identifierstyle=\color{black},
  stringstyle=\color{orange},
  morekeywords={~!+-*&^/\%},
  numbers=left,
  stepnumber=1,
}

\lstdefinestyle{customcwonumberscriptsize}{
  belowcaptionskip=1\baselineskip,
  xleftmargin=0pt,
  language=C,
  showstringspaces=false,
  basicstyle=\scriptsize\ttfamily,
  keywordstyle=\bfseries\color{purple!40!black},
  commentstyle=\itshape\color{blue},
  identifierstyle=\color{black},
  stringstyle=\color{orange},
  morekeywords={~!+-*&^/\%},
  numbers=none,
  stepnumber=2,
}

\lstdefinestyle{customcwonumbersmall}{
  belowcaptionskip=1\baselineskip,
  xleftmargin=0pt,
  language=C,
  showstringspaces=false,
  basicstyle=\small\ttfamily,
  keywordstyle=\bfseries\color{purple!40!black},
  commentstyle=\itshape\color{blue},
  identifierstyle=\color{black},
  stringstyle=\color{orange},
  morekeywords={~!+-*&^/\%},
  numbers=none,
  stepnumber=2,
}

\lstdefinestyle{customcwonumber}{
  belowcaptionskip=1\baselineskip,
  xleftmargin=\parindent,
  language=C,
  showstringspaces=false,
  basicstyle=\scriptsize\ttfamily,
  keywordstyle=\bfseries\color{purple!40!black},
  commentstyle=\itshape\color{blue},
  identifierstyle=\color{black},
  stringstyle=\color{orange},
  morekeywords={~!+-*&^/\%},
  numbers=none,
  stepnumber=2,
}

\lstdefinestyle{customcppwonumber}{
  belowcaptionskip=1\baselineskip,
  xleftmargin=\parindent,
  language=C++,
  showstringspaces=false,
  basicstyle=\Large\ttfamily,
  keywordstyle=\bfseries\color{purple!40!black},
  commentstyle=\itshape\color{blue},
  identifierstyle=\color{black},
  stringstyle=\color{orange},
  morekeywords={~!+-*&^/\%},
  numbers=none,
  stepnumber=2,
}

\usetikzlibrary{shadows}          
\newmdenv[
    tikzsetting= {fill=blueish},
    skipabove=0.33em,
    skipbelow=0.33em,
    linewidth=1pt,
    innerleftmargin=4pt,
    innerrightmargin=4pt,
    innertopmargin=2pt,
    innerbottommargin=2pt,
    linecolor=gray95,
    roundcorner=2pt, 
    shadow=true,
    shadowsize=4pt,
    shadowcolor=gray95
]{questionbox}

\newmdenv[
    tikzsetting= {fill=greenish},
    skipabove=0.33em,
    skipbelow=0.33em,
    linewidth=1pt,
    innerleftmargin=4pt,
    innerrightmargin=4pt,
    innertopmargin=2pt,
    innerbottommargin=2pt,
    linecolor=gray95,
    roundcorner=2pt, 
    shadow=true,
    shadowsize=4pt,
    shadowcolor=gray95
]{answerbox}

\newmdenv[
    skipabove=0.33em,
    skipbelow=0.33em,
    innerleftmargin=4pt,
    innerrightmargin=4pt,
    innertopmargin=2pt,
    innerbottommargin=2pt,
]{lessonbox}

\newenvironment{result}
{\begin{answerbox}}
{\end{answerbox}}

%% file: includes/macros.tex
\newcommand{\etal}{\hbox{\emph{et al.}}\xspace}
\newcommand{\eg}{\hbox{\emph{e.g.,}}\xspace}
\newcommand{\ie}{\hbox{\emph{i.e.,}}\xspace}

\newcommand{\transcoder}{TransCoder\xspace}

\newcommand{\linecode}[1]{\lstinline[escapechar=@,basicstyle=\small\ttfamily]{#1}~}

\newcounter{hypothesis}
\newcommand{\hyp}[1]{%
    \begin{result}
        \small
        \refstepcounter{hypothesis} 
        \label{hyp:\arabic{hypothesis}}
        \noindent{\underline{\bf Hypothesis \arabic{hypothesis}}.~#1}
    \end{result}
}

%% file: includes/pygments.tex
\definecolor{blueish}{RGB}{250, 250, 255}
\definecolor{greenish}{RGB}{250, 255, 250}
\definecolor{redish}{RGB}{255, 200, 200}
\definecolor{highlight}{RGB}{175, 255, 100}
\definecolor{darkred}{RGB}{139, 0, 0}
\definecolor{gray95}{gray}{0.05}

%% file: body/0.abstract.tex
\begin{abstract}

With the advent of new and advanced programming languages, it becomes imperative to migrate legacy software to new programming languages. 
Unsupervised Machine Learning-based Program Translation could play an essential role in such migration, even without a sufficiently sizeable reliable corpus of parallel source code. 
However, these translators are far from perfect due to their statistical nature. 
This work investigates unsupervised program translators and where and why they fail. 
With in-depth error analysis of such failures, we have identified that the cases where such translators fail follow a few particular patterns. 
With this insight, we develop a rule-based program mutation engine, which pre-processes the input code if the input follows specific patterns and post-process the output if the output follows certain patterns. 
We show that our code processing tool, in conjunction with the program translator, can form a hybrid program translator and  significantly improve the state-of-the-art. 
In the future, we envision an end-to-end program translation tool where programming domain knowledge can be embedded into an ML-based translation pipeline using pre- and post-processing steps. 
\end{abstract}

%% file: body/1.introduction.tex
\section{Introduction}
\label{sec:intro}

In today's software development ecosystem, Programming Languages (PL) are evolving rapidly, either as new languages or new features of existing languages.
In the past few years, many languages such as Go, Rust, Swift, TypeScript, Python3, etc.  have become popular.
It is often challenging to keep pace with such evolution\textemdash developers trained in one programming language find it hard to adapt to the new paradigm~\cite{meyerovich2013empirical}. 

There exists a large body of legacy software written in old languages like COBOL, Fortran, etc. Maintaining them is challenging as present-day developers would need to have a good understanding of these outdated languages~\cite {kizior2000does, stern2007cobol, sneed2010migrating, pu2007using, wilde2001case}. 
Organizations have been investing a lot to migrate their legacy code to newer programming languages. 
For example, in 2012, the Commonwealth Bank of Australia spent 1 Billion Australian Dollars over the subsequent five years to migrate its core banking platform \footnote{\href{https://www.reuters.com/article/us-usa-banks-cobol/banks-scramble-to-fix-old-systems-as-it-cowboys-ride-into-sunset-idUSKBN17C0D8}{https://www.reuters.com/article/us-usa-banks-cobol/banks-scramble-to-fix-old-systems-as-it-cowboys-ride-into-sunset-idUSKBN17C0D8}}. 
The Swedish bank Nordea also started their migration in 2020. 
While such migrations to newer PLs eventually save money, the investment for the migration is potentially more costly because of PLs adhering to completely different programming philosophies (\eg object-oriented vs. functional). 

To address these issues, researchers propose automated tools to convert programs written in one high-level language (\eg Java) to another high-level language (\eg Python), commonly known as \textit{Transpiler} or \textit{Transcompiler}~\cite{kulkarni2015transpiler, roziere2020unsupervised}. 
Traditionally, Transpilers are rule-based translations~\cite{babel, kimura2018javascript, 2to3}. 
A program written in the source language is represented as an abstract syntax tree, which is then translated into the target language by handwritten rules, a.k.a templates. 
Such manual rule-driven translations are not scalable, especially in the presence of external libraries and APIs. 
Furthermore, when the two language structures are very different (\eg Functional language Haskell and Procedural Object-oriented language Java), writing conversion rules may not always be possible. 
Finally, programs generated using such manual rules often lack readability. 

To overcome these issues, researchers proposed Machine Learning (ML)-based transpilers where ML models translate between two high-level programming languages by learning the statistical alignments between the two languages~\cite{aggarwal2015using, lample2018phrase, nguyen2013lexical, oda2015learning}. 
However, getting a meaningful, aligned language corpus is challenging~\cite{ahmad2022summarize, chen2018tree}. 
To this end, Roziere \etal~\cite{roziere2020unsupervised} proposed an unsupervised learning-based approach, \transcoder, where alignments between PLs are learned through back-translation~\cite{edunov2018understanding}. 
A program source language is first translated to a target language using a forward-directional translator. 
The generated target program is then translated back to the source language using a backward-direction translator. 
With joint optimization, these forward-backward translator pairs learn the alignments between the source and target languages in their respective directions without requiring an explicitly aligned corpus.  

\begin{figure*}
    \centering
    \begin{subfigure}[b]{0.49\textwidth}
         \centering
         \includegraphics[width=\textwidth]{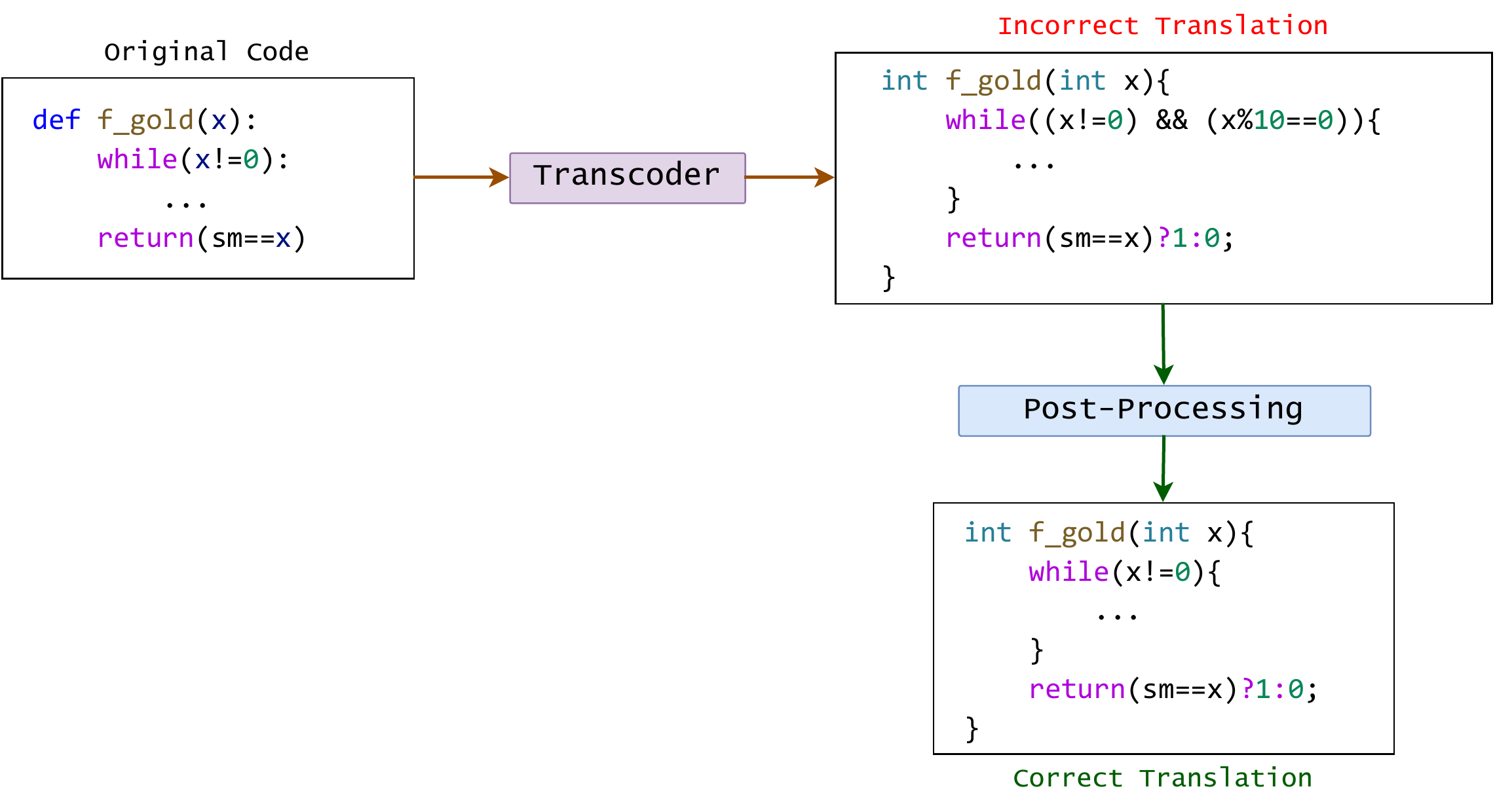}
         \caption{\small{\textbf{Post-Processing:} The \transcoder generated code has an extra incorrect \linecode{x\%10==0}condition; post-processing removed that.}}
         \label{fig:post}
    \end{subfigure}
    ~
    \begin{subfigure}[b]{0.49\textwidth}
         \centering
         \includegraphics[width=\textwidth]{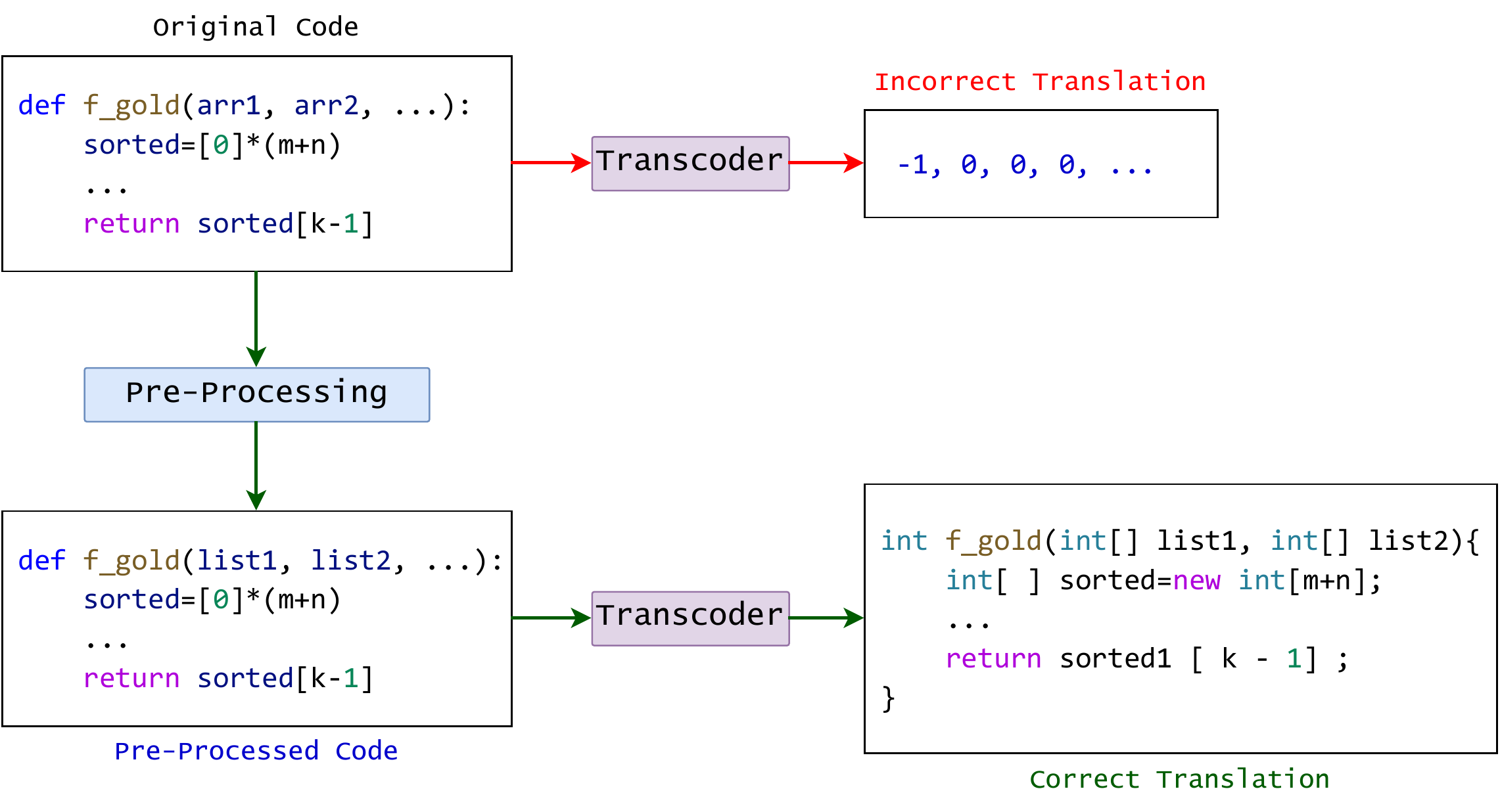}
         \caption{\small{\textbf{Pre-Processing:} \transcoder cannot translate a Python \linecode{array} parameter correctly.  When pre-processing converts the \linecode{arr} variable to \linecode{list}, the \transcoder translates correctly.}}
         \label{fig:pre}
    \end{subfigure}
    \hfill
    \caption{\textbf{\small{Motivating Example: Python to Java Translation.}}}
    \label{fig:motiv}
\end{figure*}

It turns out that unsupervised learning can outperform all the previous approaches.
However, since the \transcoder-based model is entirely driven by the statistical properties of the languages, it cannot guarantee the syntactical or semantic accuracy of the generated code. 
~\Cref{fig:motiv} shows a motivating example. While the \transcoder model {\em almost correctly} translated the input code in \Cref{fig:post}, the translated Java method contains an additional conditional clause, \linecode{x \% 10 == 0}. 
A knowledgeable developer can further mutate almost correctly translated code to obtain greater accuracy,  especially if common patterns of mistakes the model makes can be identified.

\hyp{While ``unsupervised'' translators are not perfect, their results can be post-processed if we know the model's common patterns of mistakes (\ie  ``blind spots'').} 
 
In addition, since these models are trained in an ad hoc, unsupervised way, they do not explicitly learn the syntactic and semantic alignments across language components. 
For instance, the \linecode{while}loop is semantically equivalent in Java  and Python. 
However, \linecode{for}loops in these two languages are semantically different---Java \linecode{for}loop construction often contains an updated expression for updating the loop control variable; in Python's \linecode{for}loop, such capacity is limited. Thus, \transcoder often fails to translate a Java \linecode{for}loop to a Python one. 


  \hyp{Once we identify the model's inabilities, we can systematically mutate the input code to bypass the common error-producing patterns.} 

In this pilot study, we aim to understand the common pitfalls of \transcoder and how we can improve them. 
For this purpose, we chose a large open-source unsupervised program translation model,  \transcoder, released by FaceBook AI~\cite{roziere2020unsupervised},  
which is trained on 128M GitHub repositories and has recently gained much attention. 
We then performed a rigorous manual study to find common areas where \transcoder fails to translate correctly. 
We categorize such failures into two distinct categories -- (a) semantic errors and (b) syntactic errors. 
With further investigation into each of these categories, we observe that  translations prone to semantic errors follow specific human-observable patterns and are amenable to easy post-processing corroborating hypothesis~\ref{hyp:1} (see \Cref{fig:post} as an example).
In contrast, when models make a syntactically invalid translation, we observe that the inputs follow a few specific patterns and are fixable with input program transformation through pre-processing (hypothesis~\ref{hyp:2}). 
\Cref{fig:pre} shows an example. 

ML-based code translation models come with enormous promises. However, without syntactic or semantic guidance, we cannot exploit their full potential. 
As a proof-of-concept, we incorporate such guidance with a {\em rule-based} transformer that can pre-process and/or post-process the source code; these transformers can be coupled with \transcoder to build a hybrid program translator, a.k.a. transpiler. Our initial prototype can improve the vanilla ML-based \transcoder by 86\% for Java to Python translation and 50\% for Python to Java translation. 
This indicates that guiding the ML model with program-property-aware techniques has significant potential in program translation. 

%% file: body/4.method.tex
\section{Study Design}
\label{sec:method}

\transcoder is a state-of-the-art and  popular model that accomplishes programming language translation using unsupervised learning fueled by a GeeksforGeeks unlabelled dataset. 
It is a gigantic transformer-based model trained on a public Github corpus repository of roughly 2.8 million open-source repositories. Yet, the reported accuracy is still suboptimal; \transcoder's performance is evaluated via a metric known as computational accuracy, or the ability of a translated program to produce the same output as the source code when run. The computational accuracy of \transcoder's Python to Java translation is 68.7\%, and 56.1\% the other way around.

To understand what kind of errors \transcoder commonly makes, we dug deeper into the \transcoder-generated translations using 100 examples. 
Two of the authors went through the code examples and noted their findings which were verified by another two authors. 
For each case, all of the authors reached a consensus about the type of potential error. 
To this end, we identify some common error patterns \transcoder is making. 
Leveraging these findings, we propose a hybrid technique combining machine learning and traditional rule-based solutions that can give an end-to-end solution to the code translation problem. 

\noindent
\textbf{Dataset.} Facebook AI’s Github page ~\cite{facebookresearch} provided extensive testing data for the \transcoder model taken from the GeeksForGeeks dataset. The testing dataset provided is comprised of around 280 files each in Python, Java, and C++. Each file has a method, f\_gold(), which is to be translated, along with a main method containing test cases. We randomly sampled 50 test cases for both Java to Python and Python to Java translation analysis. For each test case, we used the \transcoder to translate each file, analyzed the progress of each translation, and marked what errors were similar in multiple file translations and potential solutions.

%% file: body/5.results.tex
\section{Preliminary Results}
\label{sec:results}

\begin{table}[h]
\small
\centering
    \caption{\textbf{\small{Common Error Patterns found in TransCoder}}}
    \label{tab:error}
    \resizebox{.98\columnwidth}{!}{
\begin{tabular}{ l|r|r }
\toprule
 & \textbf{Java to Python} & \textbf{Python to Java} \\ 
&  (J2P) & (P2J) \\\midrule
1. Additional Context & 18\% & 38\%\\ 
2. Loop Conversion & 12\% & 0\%\\ 
3. Type Sensitivity & 38\% & 4\%\\ 
4. Extra Constraints & 0\% & 50\%\\ 
5. Miscellaneous Errors & 14\% & 16\%\\ 
\midrule
(Mostly) Correct & 22\% & 18\%\\ 
\bottomrule
\end{tabular}
}
\end{table}

\begin{figure*}[t]
    \centering
    \includegraphics[width=\textwidth]{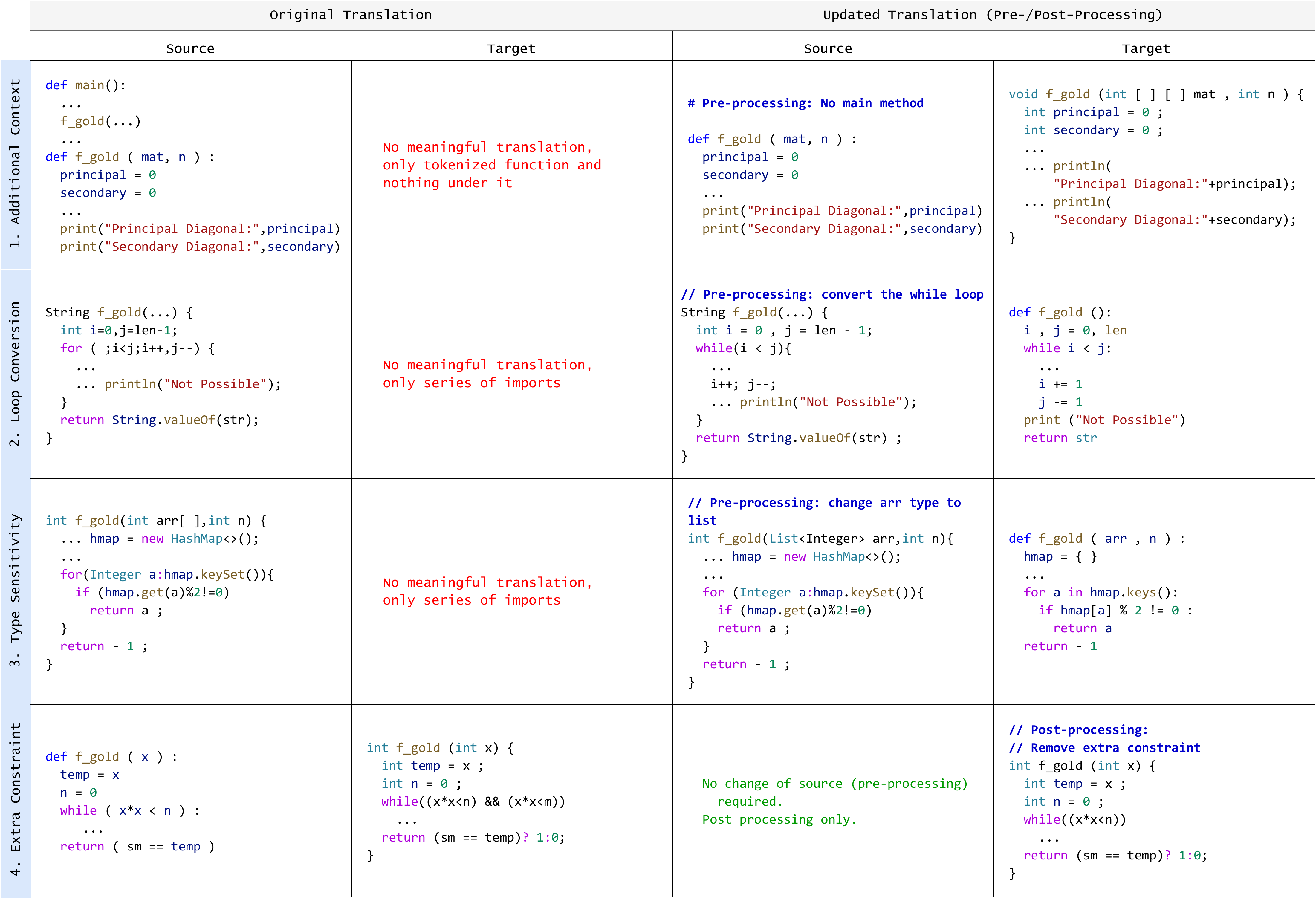}
    \caption{\textbf{\small{Detected error patterns and their proposed rule-based solutions}}}
    \label{fig:error_patterns}
\end{figure*}

Based on this study, we identify 4 different categories of errors.
~\Cref{tab:error} shows the distribution. 
In comparison to Java to Python translation (J2P), Python to Java (P2J) 
has a slightly higher rate of success\textemdash 22\% vs.~18\%. 
This section discusses the common error patterns  and potential ways to fix them using template-based pre-processing and post-processing approaches. These percentages are calculated by taking the percentage of the 50 test cases 
 in both J2P and P2J that display the mentioned errors. ~\Cref{fig:error_patterns} illustrate the errors and plausible solutions, and the errors are described in greater detail below. 

\noindent
\textbf{1. Additional Context.} The goal of the model is to accurately translate one method, typically called the \textit{focal method}. However, the focal method is often surrounded by a \texttt{main} method and test cases. We call these extra surroundings 'additional context'. 
 \transcoder tends to get confused between arguments inside and outside the method, and will sometimes translate the additional context as well, resulting in incorrect or unreadable code. 9 out of 50 Java to Python (J2P) and 19 out of 50 Python to Java (P2J) examples suffer from this problem. 

\noindent
\textit{Fixes.} Once these focal methods are translated in isolation (without the additional context), the \transcoder generates the correct output. ~\Cref{fig:error_patterns} Row 1 shows an example. While focal method \texttt{f\_gold} is called and the \texttt{main} method is still in the context, \transcoder could not generate any meaningful translation. However, when we remove the additional contexts, the translation accuracy significantly improves.  

In the rest of the paper, we treat the \transcoder as a function translator. The translation errors observed will henceforth be mainly errors that occurred when we singularly translated the functions using the \transcoder. 



\noindent
\textbf{2. Loop Conversion.} 
Vanilla \transcoder performs poorly while translating complex \linecode{for}loop to \linecode{while}loop, especially for Java to Python translation. 
As Java \linecode{for}loops generally allow more functionalities than Python \linecode{for}loops (\eg different increment of the loop variables, more variables, more conditions), 
the \transcoder model has difficulty translating complex \linecode{for}loops from Java to Python.
Complex for loops appeared in 6 out of 50 samples, and all of them could not produce correct outputs, where 4 out of the 6 produced garbage translation.

\noindent
\textit{Fixes.} 
We hypothesize that it would be beneficial to convert the \linecode{for}loops to \linecode{while}loops before passing the input to \transcoder, as the latter is syntactically equivalent in Python and Java. Thus, as a pre-processing step, we performed semantic preserving transformation to covert \linecode{for} to \linecode{while}. 
Such pre-processing significantly improved the translation of all 6 incorrect cases. ~\Cref{fig:error_patterns} second row shows an example.

\noindent
\textbf{3.~Type Sensitivity.} 
We find that \transcoder can be sensitive to certain types. For example, 
19 out of 50 examples J2P examples 
contain an array as a parameter. 
\transcoder fails to translate all these cases, as shown in the third-row of~\Cref{fig:error_patterns}. For P2J as well, (see~\Cref{fig:pre}), when the input focal method contains two or more parameters with names \linecode{arr}, \transcoder fails to translate them. Note that, since Python is a dynamically typed language, we have to rely on the variable names to infer their types. However, the corresponding ground truth Java code confirms the intended type is indeed an array. 

\noindent
\textit{Fixes.} We explore a preprocessing step where without changing the code's semantics we tried to use equivalent types or classes. For instance, in the above case, we change all the array parameter references in the Java code to a List of the equivalent data type, as the Python translation of a Java array versus a Java List is identical. Note that we can not use the exact same data type when converting an array to List. Instead, we must use the wrapper class data type (int to Integer, double to Double, etc).
Such type transformation in the pre-processing helped us to improve \transcoder's performances across all the 19 cases.

\noindent
\textbf{4.~Generating Extra Constraints.} 
	The most prominent issue for Python to Java translation is generating extraneous logical operators to \linecode{if} and \linecode{else if} and \linecode{while} statements. Out of 50 examples, 25 had such issues. Although such additional logical operators are syntactically valid, they can potentially change code semantics. The last row of ~\Cref{fig:error_patterns} is an example. 	

\noindent
\textit{Fixes.} As a post-processing step, we discard all the logical constraints that do not appear in the source version. This is due to the observation that although the model appends logical constraints, it never modifies the original conditions.

\noindent
\textbf{Overall Results.} 
The performance of each mutation is measured by the rate of success. We classify "success" in two cases: 

    1. If a translated program does not compile, a \textit{success} is when the translation of the program after applying mutations compiles.  
    
    2. If a translated program does compile, but with error, a \textit{success} is when the translation of the program after applying mutations runs more similarly to the original program. More specifically, if the translated code can be more easily interpreted to have the same functionality as the source code, we would classify the mutation as a success.
    
To evaluate the effectiveness of each mutation, we first determined of which sampled test cases each mutation was applicable to. After translating the both original source code and the mutated source code, we classified the mutation as a success or fail for each test case. If multiple mutations were applicable to any test case, we would apply all possible combinations to ensure successes in each translation. The rate of success of a specific mutation, or rule, is computed as the number of successes divided by the number of cases it was applicable to. Each of our identified mutations have a 100\% success rate, though there are errors for which we have not yet discovered a viable mutation for yet (Miscellaneous Errors in Table 1). 



%% file: body/6.related.tex
\section{Related Works}
\label{sec:related}
Multiple previous studies have investigated the possibility of programming language translation through machine learning. 
However, almost all studies rely on supervised learning~\cite{chen2018tree, ahmad2021plbart, msr2021codexglue, chakraborty2022natgen, feng2020codebert, guo2021graphcodebert}. 
This approach is unrealistic, though accurate, as it is difficult to accumulate a high total of labeled, correctly translated, datasets~\cite{chen2018tree}.  

While it is difficult to come across labeled datasets, some researchers have found it effective to train their model based on a technique called back-translation~\cite{roziere2020unsupervised, ahmad2022summarize}. 
Being unsupervised, the capability of these models are not limited by the quantity of the annotated parallel data, making them state-of-the-art for program translation. 
In this work, we case study one such model, \transcoder~\cite{roziere2020unsupervised}. 
Other research has delved into the application of SMT (statistical machine translation)~\cite{nguyen2013lexical, karaivanov2014phrase, aggarwal2015using} models in the translation of programming languages. These studies have also reached conclusions similar to this project, that a majority of test cases have errors, but only need small fixes to produce correct translations. These models can also be improved in a more program-analysis-oriented approach, as our techniques demonstrate as well~\cite{nguyen2013lexical}. 

Researchers also proposed translation models for in-language code transformation for syntactic repair~\cite{ahmed2021synfix, ahmed2022synshine}, semantic program repair~\cite{Chakraborty2020codit, chen2019sequencer, tufano2019empirical}, refactoring~\cite{aniche2020effectiveness, sheneamer2020automatic}, etc. 
More recently, researchers have been proposing general purpose code transformation models ``pre-trained'' from developer-written code transformation collected from GitHub~\cite{zhang2022coditt5}, or rule-based transformations~\cite{chakraborty2022natgen}. 
In the future, we aim at investigating both the syntactic and semantic repair models as our pre-processing and post-processing components.  

%% file: body/7.future-work.tex
\section{Conclusion \& Future Work}
\label{sec:conclusion}

\underline{\it Paper Summary.} In this paper, we discuss the pitfalls of unsupervised program translators and present the potential of program-property-aware rules that can guide the ML-based translation as pre-/post- processing steps. We developed a proof-of-concept in-language program transformer for pre-processing the input and post-processing the output of \transcoder. We show that a simple rule-based in-language program transformer can significantly outperform program translation performance. Our preliminary results, along with detailed instructions to replicate each mutation, are publicly available at {\href{https://github.com/kzh23/Replication-Package-ICSE-NIER-2023-Unsupervised-ML}{https://github.com/kzh23/Replication-Package-ICSE-NIER-2023-Unsupervised-ML}}.
While the ML-based translator relies on statistical knowledge embedded in ``big data'', we propose to embed programming domain knowledge into the translation pipeline. 


\underline{\it Future Work.} This paper serves as an initial attempt toward combining ML-based program translation and program analysis-based program mutation. We aim to build more sophisticated and automated techniques for program transformation in the future. As evidenced by our initial results, guiding the ML-based tools with program-property-aware rules has immense potential in program translation. In the future, we will leverage how smartly incorporate such guidance in the ML pipelines. For instance, currently, the vanilla \transcoder can only translate methods in isolation. Such limitations will hinder the adaptability of the proposed techniques in real life, where an entire project written in legacy language needs to be translated. We will further study the applicability of the proposed technique in low-resourced languages where we will not get enough sample data for the training ML model; in such cases, the rule-based approach may need to provide more guidance. 

 To this end, we envision building a scalable, modular, end-to-end system combining pre-processing, translation, and post-processing steps. We also intend to investigate the usage of code editing models~\cite{zhang2022coditt5, chakraborty2022natgen} as pre-processing and program repair tools~\cite{chen2019sequencer, ye2022neural, yasunaga2021break} as post-processing steps for better generalization. 